\documentclass[twocolumn,pra,aps]{revtex4}
\usepackage{graphicx}

\def\duzomniejsze{<\kern-.7mm<}
\def\duzowieksze{>\kern-.7mm>}

\def\textbf#1{{\bf #1}}
\def\beq{\begin{equation}}
\def\eeq{\end{equation}}
\def\be{\begin{equation}}
\def\ee{\end{equation}}
\def\ben{\begin{eqnarray}}
\def\een{\end{eqnarray}}
\def\beqa{\begin{eqnarray}}
\def\eeqa{\end{eqnarray}}
\def\eea{\end{array}}
\def\bea{\begin{array}}
\newcommand{\bei}{\begin{itemize}}
\newcommand{\eei}{\end{itemize}}
\newcommand{\bee}{\begin{enumerate}}
\newcommand{\eee}{\end{enumerate}}

\def\>{\rangle}
\def\<{\langle}

\begin{document}

\title{Simulation of continuous variable quantum games without entanglement}

\begin{abstract}
A simulation scheme of quantum version of Cournot's Duopoly is
proposed, in which there is a new Nash equilibrium that may be also
Pareto optimal without any entanglement involved. The unique
property of this simulation scheme is decoherence-free against the
symmetric photon loss. Furthermore, we analyze the effects of the
asymmetric information on this simulation scheme and investigate the
case of asymmetric game caused by asymmetric photon loss. A
second-order phase transition-like behavior of the average profits
of the firm 1 and firm 2 in Nash equilibrium can be observed with
the change of the degree of asymmetry of the information or the
degree of "virtual cooperation". It is also found that asymmetric
photon loss in this simulation scheme plays a similar role with the
asymmetric entangled states in the quantum game.

\end{abstract}
\author{Shang-Bin Li}\email{stephenli74@yahoo.com.cn}

\affiliation{Research and development department of Amertron
optoelectronic (Kunshan) Ltd., Jingde road 28, kunshan, Suzhou, P.R.
China}

\maketitle

\section * {I. INTRODUCTION}

Recently, significant interests have been focused on generalizing
the classical notion of game theory to an analogous quantum version
\cite{meyer,eisert}, the so-called quantum game theory, which is a
new born branch of quantum information theory. Quantum games and
quantum strategies can exploit both quantum superposition
\cite{meyer,Enk} and quantum entanglement \cite{eisert,Enk2002}.
Meyer has pointed the way for generalizing the classical game theory
to quantum domain by utilizing the quantum superposition
\cite{meyer}, though it has been argued that the Meyer's scheme can
be classically realized \cite{Enk}. By making use of the
entanglement resource, the Prisoner's Dilemma has been quantized
through the scheme of Eisert et al. and shown that the game ceases
to pose a dilemma if a restricted quantum strategies are allowed for
\cite{eisert,Benjamin2001prl,Flitney2007pla}, which has been
experimentally demonstrated by Du et al. \cite{Du2002}. Since then
more work has been done on quantum prisoners' dilemma
\cite{Benjamin2001,Iqbal,Iqbal2004,Cao} and a number of other games
have been generalized to the quantum realm (For review, see
Refs.\cite{Iqbal2005,Guo2008} and references therein). Besides those
games in which the players have finite number of strategies, the
quantization of classical Cournot's duopoly in which the players can
access to a continuous set of classical strategies has been also
investigated
\cite{Li2002,Du2003,Li2006,Lo2003,Lo2005,Du2005,Zhou,Qin}. Classical
Cournot's duopoly exhibits a dilemma-like situation, in which the
unique Nash equilibrium is inferior to the Pareto optimum. For the
quantum version of Cournot's duopoly, even though two players both
act as "selfishly" in the quantum game, they are found to virtually
cooperate due to the quantum entanglement between them
\cite{Li2002}. However, entanglement is usually very fragile against
the decoherence caused by the interaction with the surrounding
environment. The advantage of most of the previous quantum games are
not robust against the noise \cite{Johnson2001} and the unique
properties of various quantum games different from their classical
counterpart will disappear in the limit of decoherence
\cite{Flitney2005,Imoto}. However, Chen et al. have discussed
decoherence in quantum prisoners' dilemma \cite{Chen2003}, and found
some kinds of decoherence have no effects on the Nash equilibria in
the quantized prisoners' dilemma with maximally entangled states.
Here, we present the simulation schemes of the quantized symmetric
or asymmetric Cournot's Duopoly, in which there is not any
entanglement involved. However, the players can also escape the
dilemma-like situation. In this scheme, classical measuring
apparatus provides more profits than the quantum measuring
apparatus. In the asymmetric game, "virtual cooperation" does not
give any advantage to the weaker of two firms if the degree of the
asymmetry exceeds certain threshold value. The most significant
aspect of this scheme is its symmetric decoherence-free, namely
certain kind of symmetric decoherence does not alter the unique
property of this quantized Cournot's duopoly.

\section * {II. CONTINUOUS VARIABLE QUANTUM GAME WITHOUT ENTANGLEMENT}

To make this paper self-contained, we briefly outline the classical
Cournot's duopoly and its quantization version in Ref.\cite{Li2002}.
In a simple version of Cournot's model for the duopoly, two firms
simultaneously decide the quantities $q_1$ and $q_2$ respectively
of a homogeneous product released on the market. Suppose $Q$ is the
total quantity, i.e., $Q=q_1+q_2$, and the market-clearing price is
given by $P(Q)=a-Q$ for $Q\leq{a}$ and $P(Q)=0$ for $Q>{a}$. The
unit cost of producing the product is assumed to be $c$ with $c<a$.
The payoffs or profits of the firms can be written as \be
u_i(q_1,q_2)=q_i[P(Q)-c]=q_i[k-(q_1+q_2)], \ee where $k=a-c$ is a
constant and $i=1,2$. Solving for the Nash equilibrium (immune to
unilateral deviations) yields the cournot equilibrium, \be
q^{\ast}_1=q^{\ast}_2=\frac{k}{3}. \ee At this equilibrium the
payoff for the firm $i$ ($i=1,2$) is $k^2/9$, which is not the
optimal solution. If two firms can cooperate and restrict their
quantities to $q^{\prime}_1=q^{\prime}_2=k/4$, they can both get
higher payoff $k^2/8$, which is the highest profits they can attain
in this symmetric game. However, in a competing game, the objective
of each firm is to maximize its individual payoff, and avoid the
unilateral deviation causing decrease of its profit. This individual
rationality confines the strategies of two firms in the Nash
equilibrium point.

The quantization version of the classical Cournot's duopoly showed
that quantum entanglement creates virtual cooperation of two firms
and the larger entanglement can guarantee the players (in the Nash
equilibrium) a payoff closer to the highest feasible payoff, i.e.
the Pareto-optimal payoff. Notwithstanding the domination role of
entanglement in various quantum strategies, we attempt to present
a simulation scheme of quantized Cournot's Duopoly, in which there
is not any intermediate quantum entanglement involved. We utilize
two single-mode optical fields which are initially in the vacuum
state $|0\rangle_1\otimes|0\rangle_2$. Then, two optical
fields are sent to firm 1 and firm 2, respectively. The strategic
moves of firm 1 and firm 2 are represented by the displacement
operators $\hat{D}_1$ and $\hat{D}_2$ locally acted on their
individual optical fields. The players are restricted to choose
their strategies from the sets \be
S_i=\{\hat{D}_i(x_i)=\exp[\frac{\sqrt{2}}{2}x_i(a^{\dagger}_i-a_i)]|x_i\in[0,\infty)\},~i=1,2
\ee where $a_i$ and $a^{\dagger}_i$ are the annihilation and
creation operators of the $i$th mode optical field, respectively. In this stage,
the state of the game becomes a direct product of two coherent
states $|\frac{\sqrt{2}}{2}x_1\rangle$ and
$|\frac{\sqrt{2}}{2}x_2\rangle$, \be
|\psi\rangle=|\frac{\sqrt{2}}{2}x_1\rangle\otimes|\frac{\sqrt{2}}{2}x_2\rangle.
\ee Having executed their moves, firm 1 and firm 2 forward their
optical fields to the final measurement, prior to which a beam
splitter operation
$\hat{J}(\gamma)=\exp[i\gamma({a}^{\dagger}_1a_2+{a}^{\dagger}_2a_1)]$
($\gamma\in[0,\frac{\pi}{4})$) is carried out. Therefore the final
state prior to the measurement can be expressed as \beqa
|\Psi\rangle&=&|\frac{\sqrt{2}}{2}{x}_1\cos\gamma+\frac{\sqrt{2}}{2}i{x}_2\sin\gamma\rangle\nonumber\\
&&\otimes|\frac{\sqrt{2}}{2}{x}_2\cos\gamma+\frac{\sqrt{2}}{2}i{x}_1\sin\gamma\rangle.
\eeqa Then, a measurement on the photon number of the optical
fields is carried out, which is usually done by photon-detector.
The measurement is also one of the key issues in quantum games.
Different kinds of measurement schemes can alter the
characteristics of the games \cite{Measurement,Monty2001}. The
measurement schemes mainly depend on the measuring apparatus used.
In what follows, we analyze two kinds of measuring apparatuses and
investigate their influence on this simulation scheme: (1) "classical" measuring
apparatus which can only give out the expected values of the
photon numbers of the optical fields such as the optical power
meter. (2) quantum measuring apparatus, such as the highly
sensitive quantum photon-counter which can measure the photon
number and its distribution.

\subsection * {A. 'Classical' measuring apparatus}

Firstly, we assume only the expected values of the photon number
of the optical fields are measured by the 'classical' measuring
apparatus, and the expected values of the photon number are $n_i$
($i=1,2$), which are regarded as the individual quantities. Hence
the payoffs are given by \be
u^{Q}_i(\hat{D}_1,\hat{D}_2)=u_i(n_1,n_2), \ee where the
superscript "Q" denotes "quantum". The classical Cournot's Duopoly
can be faithfully represented when $\hat{J}(\gamma)|_{\gamma=0}=I$
(the identity operator), in which the quantities
of firm 1 and firm 2 is $n_1=x^2_1/2$ and $n_2=x^2_2/2$,
respectively. For the final state in Eq.(5), the measurement gives
the respective quantities of two firms \beqa
n_1=\frac{1}{2}(x^2_1\cos^2\gamma+x^2_2\sin^2\gamma),\nonumber\\
n_2=\frac{1}{2}(x^2_2\cos^2\gamma+x^2_1\sin^2\gamma). \eeqa Then,
the quantum profits for two firms are given by \beqa
u^{Q}_1(\hat{D}_1,\hat{D}_2)=\frac{1}{2}(x^2_1\cos^2\gamma+x^2_2\sin^2\gamma)[k-\frac{1}{2}(x^2_1+x^2_2)],\nonumber\\
u^{Q}_2(\hat{D}_1,\hat{D}_2)=\frac{1}{2}(x^2_2\cos^2\gamma+x^2_1\sin^2\gamma)[k-\frac{1}{2}(x^2_1+x^2_2)].
\eeqa Solving for the Nash equilibrium gives the unique one \be
x^{\ast}_1=x^{\ast}_2=\sqrt{\frac{2k\cos^2\gamma}{1+2\cos^2\gamma}}.
\ee The profits of two firms at this equilibrium are given by \be
u^{Q}_1=u^{Q}_2=\frac{k^2\cos^2\gamma}{(1+2\cos^2\gamma)^2}. \ee
From Eq.(10), we can see that the profit at the equilibrium
increases from the classical payoff $\frac{k^2}{9}$ to
pareto-optimal payoff $\frac{k^2}{8}$ when $\gamma$ increases in the
range of $\gamma\in[0,\frac{\pi}{4})$. Obviously, not any
intermediate quantum entanglement has been involved in this scheme.
But this fact does not impede the successful escaping from the
dilemma when $\gamma\rightarrow\frac{\pi}{4}$. It is true that
complete classical system can be used to implement the proposed
scheme. For example, by making use of the source of classical light
or electromagnetic wave, two modulators, beam splitter, and power
meters, one can realize this simulation scheme. Though there exists
the intermediate entanglement in the quantization scheme of
Ref.\cite{Li2002}, the final state in Eq.(11) in Ref.\cite{Li2002}
is also not entangled. In this sense, the present scheme and the one
in Ref.\cite{Li2002} can be regarded as the same kind but with
different definitions of quantum strategies.

\subsection * {B. Quantum measuring apparatus}

In this subsection, we assume that a quantum measuring apparatus
can be used to measure the photon number distributions of the
optical fields. The measured value of photon number is regarded as
the quantity of the respective strategy, then the average payoff
is calculated based on the probability distribution of the photon
number. For the final state in Eq.(5), the payoffs are given by
\be u^{Q}_i(\hat{D}_1,\hat{D}_2)=\langle{u_i(m_1,m_2)}\rangle, \ee
where \be
\langle{u_i(m_1,m_2)}\rangle=\sum^{\infty}_{m_1,m_2=0}u_i(m_1,m_2)P_{m_1,m_2}
\ee denotes the average of $u_i(m_1,m_2)$ taken over all possible
values of $m_1$ and $m_2$ with the Poisson distribution \beqa
P_{m_1,m_2}&=&e^{-\frac{1}{2}(x^2_1+x^2_2)}\frac{(\frac{1}{2}(x^2_1\cos^2\gamma+x^2_2\sin^2\gamma))^{m_1}}{m_1!}\nonumber\\
&&\frac{(\frac{1}{2}(x^2_2\cos^2\gamma+x^2_1\sin^2\gamma))^{m_2}}{m_2!}.
\eeqa For simplicity, we assume $a$ and $c$ tend to infinity but
keeping $k=a-c\geq1$ a finite constant. In this case, the quantum
payoffs for two firms are given by \beqa
u^{Q}_1(\hat{D}_1,\hat{D}_2)=\frac{1}{2}(x^2_1\cos^2\gamma+x^2_2\sin^2\gamma)[k-1-\frac{1}{2}(x^2_1+x^2_2)],\nonumber\\
u^{Q}_2(\hat{D}_1,\hat{D}_2)=\frac{1}{2}(x^2_2\cos^2\gamma+x^2_1\sin^2\gamma)[k-1-\frac{1}{2}(x^2_1+x^2_2)].\nonumber\\
\eeqa In this case, when $\hat{J}(\gamma)=I$ (the identity
operator), the scheme can not return to the classical Cournot's
Duopoly. Comparing the payoffs in Eq.(14) and Eq.(8), we can find
that quantum fluctuation causes the reduce of the payoffs. Solving
for the Nash equilibrium yields the unique one \be
x^{\ast}_1=x^{\ast}_2=\sqrt{\frac{2(k-1)\cos^2\gamma}{1+2\cos^2\gamma}}.
\ee The profits of two firms at this equilibrium are given by \be
u^{Q}_1=u^{Q}_2=\frac{(k-1)^2\cos^2\gamma}{(1+2\cos^2\gamma)^2}. \ee
From Eq.(16), we can see that the profit at the equilibrium
increases from the $\frac{(k-1)^2}{9}$ to $\frac{(k-1)^2}{8}$ when
$\gamma$ increases from $0$ to $\frac{\pi}{4}$. The above results
show the scheme using classical measuring apparatus has advantage to
the scheme using the quantum measuring apparatus.

Here, for avoiding the emergence of the situation $m_1+m_2>a$ with
nonzero probability, after Eq.(13) it has been assumed $a$ and $c$
tend to infinite but keeping $k=a-c$ a finite constant, which
guarantees the payoff summed over $m_1$, $m_2$ from $0$ to infinity
in Eq.(12) has the analytical expression in Eq.(14). For very large
but finite value of $a$ and $c$, the probability $P_{m_1,m_2}$ in
Eq.(13) corresponding to $m_1+m_2>a$ tends to very very small, and
the summation of those terms with $m_1+m_2>a$ in Eq.(12) should be
written as the summation of $-cm_1P_{m_1,m_2}$ or $-cm_2P_{m_1,m_2}$
which also tends to small enough to guarantee the payoff well
approximated by the Eq.(14). However, for other cases with small
values of $a$, the situation becomes very complicate. The payoff in
Eq.(14) is not valid and need to be revised. It is very difficult to
obtain an analytical result in this situation. Our numerical results
show the optimal payoff of two firms is between $k^2/8$ and
$(k-1)^2/8$ as $\gamma$ tends to $\pi/4$. For examples, in the case
with $a=6$, $c=1$, $k=5$, $\gamma=\pi/4$, the optimal payoff is
about $2.02487$, which is near $(k-1)^2/8=2$; in the case with
$a=10$, $c=5$, $k=5$, $\gamma=\pi/4$, the optimal payoff is about
$2.00006$. Thus it is conjectured that the conclusion that scheme
using classical measuring apparatus has advantage to the scheme
using the quantum measuring  apparatus is valid even in the cases
with small values of $a$ and $c$.

\section * {III. SYMMETRIC DECOHERENCE-FREE ASPECTS OF THIS QUANTIZED SCHEME}

The most significant characteristics of our simulation scheme is its
symmetric decoherence-free. Previous works have shown that the
decoherence could destroy the advantage of quantum game
\cite{Johnson2001,Flitney2005,Imoto}. In our simulation scheme, the
advantage of quantum game is robust against the symmetric
photon-loss, where the decoherence caused by the photon loss can be
described by the following master equation \cite{Phoenix1990}, \be
\frac{\partial\rho(t)}{\partial{t}}=\sum^{2}_{i=1}\kappa\hat{a}_i\rho(t)\hat{a}^{\dagger}_i-
\frac{\kappa}{2}(\hat{a}^{\dagger}_i\hat{a}_i\rho(t)+\rho(t)\hat{a}^{\dagger}_i\hat{a}_i),
\ee where $\kappa$ is the decay rate. $\rho(t)$ represents the whole
state of the firm 1 and firm 2. The evolving state can be expressed
as, \be
|\frac{\sqrt{2}}{2}x_1(t)\rangle_1|\frac{\sqrt{2}}{2}x_2(t)\rangle_2
\ee where
$|\frac{\sqrt{2}}{2}x_i(t)\rangle_i=|\frac{\sqrt{2}}{2}x_ie^{-\frac{\kappa}{2}t}\rangle_i$
($i=1,2$). Then forward the evolving state into the beam splitter
and we have \beqa
|\Psi(t)\rangle&=&|\frac{\sqrt{2}}{2}{x}_1e^{-\frac{\kappa}{2}t}\cos\gamma+\frac{\sqrt{2}}{2}i{x}_2e^{-\frac{\kappa}{2}t}\sin\gamma\rangle\nonumber\\
&&\otimes|\frac{\sqrt{2}}{2}{x}_2e^{-\frac{\kappa}{2}t}\cos\gamma+\frac{\sqrt{2}}{2}i{x}_1e^{-\frac{\kappa}{2}t}\sin\gamma\rangle.
\eeqa From the above quantum state, we can immediately obtain the
following conclusion: If both firms have the complete information
about the photon loss, they can adjust their strategies according to
the transformation $x_i\rightarrow{x}_ie^{\frac{\kappa}{2}t}$, which
can guarantee the final payoffs are invariant under the influence of
the photon loss process.

\section * {IV. ASYMMETRIC INFORMATION ASPECTS OF THIS SIMULATION SCHEME}

\begin{figure}
\centerline{\includegraphics[width=2.5in]{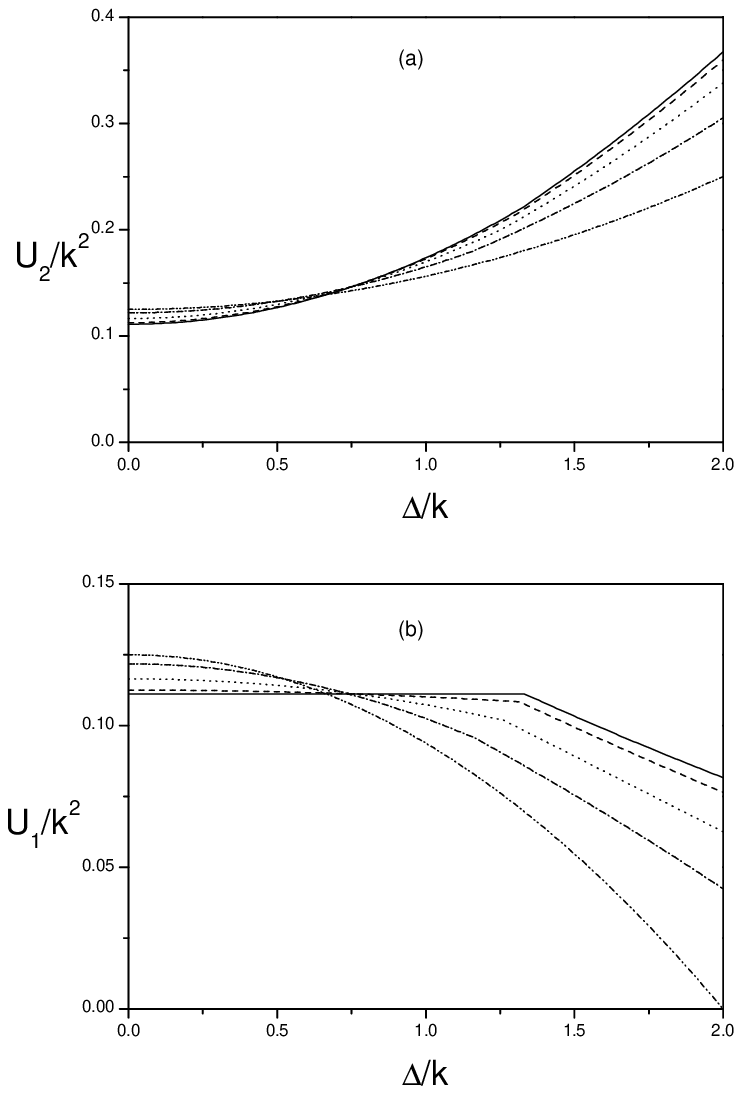}}
\caption{(a) The scaled payoff $U_2/k^2$ of the firm 2 and (b) the
scaled payoff $U_1/k^2$ of the firm 1 at Nash equilibrium are
plotted as the function of $\Delta/k$ for different values of
$\gamma$ with $\theta=0.5$. (Solid line) $\gamma=0$; (Dash line)
$\gamma=\pi/16$; (Dot line) $\gamma=\pi/8$; (Dash dot line)
$\gamma=3\pi/16$; (Dash dot dot line)
$\gamma=(\frac{\pi}{4})^{-}$. }
\end{figure}
\begin{figure}
\centerline{\includegraphics[width=2.5in]{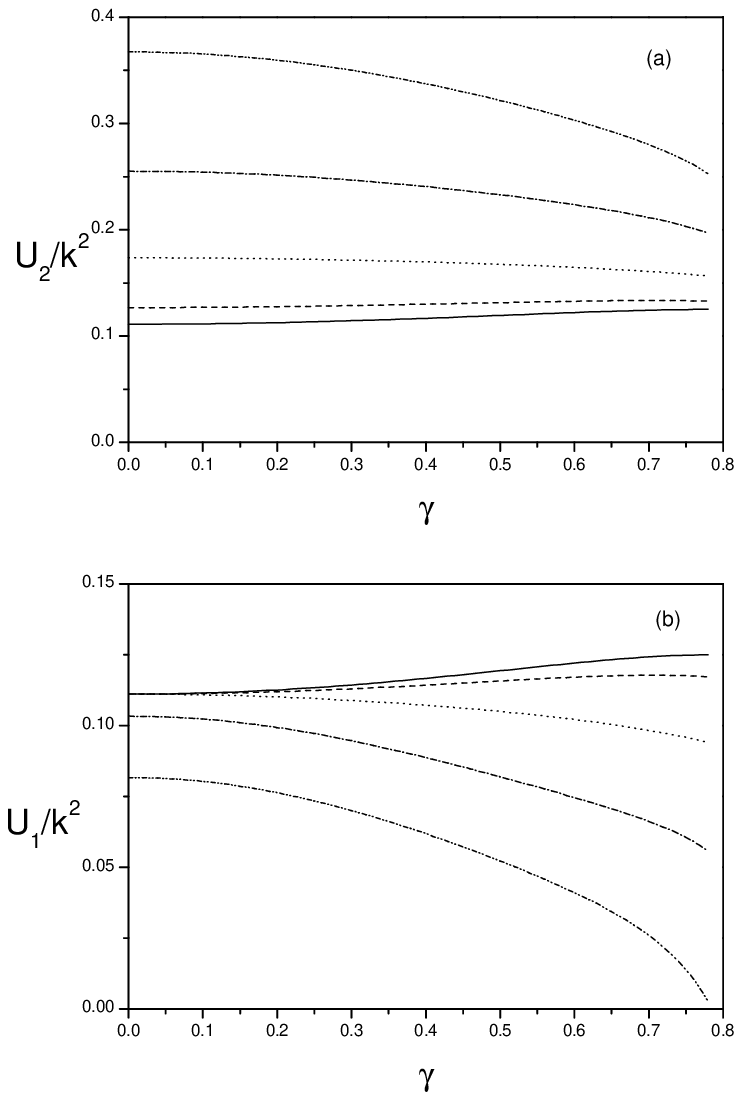}}
\caption{(a) The scaled payoff $U_2/k^2$ of the firm 2 and (b) the
scaled payoff $U_1/k^2$ of the firm 1 at Nash equilibrium are
plotted as the function of $\gamma$ for different values of
$\Delta/k$ with $\theta=0.5$. (Solid line) $\Delta/k=0$; (Dash line)
$\Delta/k=0.5$; (Dot line) $\Delta/k=1$; (Dash dot line)
$\Delta/k=1.5$; (Dash dot dot line) $\Delta/k=2^-$. }
\end{figure}

Recently, the quantum game with asymmetric information has been
investigated and some novel phenomenons caused by asymmetry of
information have been revealed \cite{Du2003}. It is very interesting
to study how the asymmetric information can alter the aspects of the
present simulation scheme of the original quantized game. In the
case with asymmetric information, firm 1 does not know what $c_2$
(firm 2's unit cost) is, only knows that $c_2=c_H$ with probability
$\theta$ and $c_2=c_L$ with probability $1-\theta$ ($c_H>c_L$). Yet
firm 2 knows with certainty the unit cost $c_2$ of its product as
well as that of firm 1's ($c_1$). So, the Eq.(8) should be replaced
by \beqa
u^{Q}_1(\hat{D}_1,\hat{D}_2)&=&\frac{1}{2}(x^2_1\cos^2\gamma+x^2_2\sin^2\gamma)[a-c_1-\frac{1}{2}(x^2_1+x^2_2)],\nonumber\\
u^{Q}_{2H}(\hat{D}_1,\hat{D}_2)&=&\frac{1}{2}(x^2_2\cos^2\gamma+x^2_1\sin^2\gamma)[a-c_H-\frac{1}{2}(x^2_1+x^2_2)],\nonumber\\
u^{Q}_{2L}(\hat{D}_1,\hat{D}_2)&=&\frac{1}{2}(x^2_2\cos^2\gamma+x^2_1\sin^2\gamma)[a-c_L-\frac{1}{2}(x^2_1+x^2_2)].\nonumber\\
\eeqa For convenience we denote the strategy by $x_i$ when it is
$\hat{D}_i(x_i)$. Let $\{x^{\ast}_1,x^{\ast}_{2H},x^{\ast}_{2L}\}$
be the Bayes-Nash equilibrium. Then $x_2=x^{\ast}_{2H(L)}$ is chosen
to maximize $u^{Q}_{2H(L)}(x^{\ast}_1,x_2)$ and $x_1=x^{\ast}_1$ is
chosen to maximize
$\theta{u}^{Q}_{1}(x_1,x^{\ast}_{2H})+(1-\theta){u}^{Q}_{1}(x_1,x^{\ast}_{2L})$.
Solving the three optimization problem yields the Bayes-Nash
equilibrium. For simplicity, we assume that
$c_1=\theta{c_H}+(1-\theta)c_L$, $k=a-c_1$, $\Delta=c_H-c_L$. After
calculation, the unique Bayes-Nash equilibrium can be obtained \beqa
x^{\ast{2}}_1&=&\frac{2k\cos^2\gamma}{1+2\cos^2\gamma},\nonumber\\
x^{\ast{2}}_{2H}&=&\frac{2k\cos^2\gamma-4\cos^4\gamma(a-c_{H})-(1-\theta)\Delta}{1-4\cos^4\gamma},\nonumber\\
x^{\ast{2}}_{2L}&=&\frac{2k\cos^2\gamma-4\cos^4\gamma(a-c_{L})+\theta\Delta}{1-4\cos^4\gamma}.
\eeqa In the above derivation, it has been assumed that
$\max[\frac{2(c_H-c_1)-k}{k-(c_H-c_1)},0]<\cos(2\gamma)$. When
$\frac{2(c_H-c_1)-k}{k-(c_H-c_1)}\geq\cos(2\gamma)>0$, the unique
Bayes-Nash equilibrium can be obtained \beqa
x^{\ast{2}}_1&=&\frac{2\cos^2(\gamma)[\theta{k}-\theta(1-\theta)\Delta+k\cos(2\gamma)]}{\theta+\cos(2\gamma)[2+\cos(2\gamma)]},\nonumber\\
x^{\ast{2}}_{2H}&=&0,\nonumber\\
x^{\ast{2}}_{2L}&=&\frac{2\cos^2(\gamma)[\theta\Delta+(k+\theta\Delta)\cos(2\gamma)]}{\theta+\cos(2\gamma)[2+\cos(2\gamma)]}.
\eeqa In the following, we consider the iterative game. When
$\frac{2(c_H-c_1)-k}{k-(c_H-c_1)}\geq\cos(2\gamma)>0$, the average
profits in the iterative game are given by \beqa
\bar{U}^{Q}_1&=&[\cos^2\gamma(4\theta((k + \Delta(\theta - 1))\theta
+k\cos2\gamma)\nonumber\\
      &&\cdot(-2\Delta\theta^2 + 2k\theta +
      2\Delta\theta + k - 2(k(\theta - 3) \nonumber\\
      &&+\Delta(\theta - 1)\theta)\cos2\gamma + k\cos4\gamma)
     \cos^2\gamma \nonumber\\
     &&+ (\theta - 1)
     (2(k(\theta - 2) + \Delta\theta(\theta + 1))\cos2\gamma \nonumber\\
     &&+
      \theta(-2k + \Delta + 2\Delta\theta +
      \Delta\cos4\gamma))\nonumber\\
     &&\cdot(2(\Delta(\theta - 1)\theta + k(\theta + 2))\cos2\gamma \nonumber\\
     &&+
      \theta(2k - \Delta + 2\Delta\theta - \Delta\cos4\gamma)))]\nonumber\\
      &&/
  [4(2\theta + 4\cos2\gamma + \cos4\gamma + 1)^2]\nonumber\\
\bar{U}^{Q}_2&=&[\cos^2\gamma(4\theta((k + \Delta(\theta - 1))\theta
+
      k\cos2\gamma)\nonumber\\
      &&\cdot(2\Delta\theta^2 + 2k\theta + k -
      2\Delta - 2(k(\theta - 3) \nonumber\\
      &&+ \Delta(\theta^2 - 5\theta +
          4))\cos2\gamma \nonumber\\
          &&+ (k + 2\Delta(\theta - 1))
       \cos4\gamma)\sin^2\gamma \nonumber\\
       &&- (\theta - 1)
     (\theta(2k + \Delta + 2\Delta\theta + \Delta\cos4\gamma) \nonumber\\
     &&-
       2(k(\theta - 2) + \Delta(\theta - 3)\theta)\cos2\gamma)^
      2)]\nonumber\\
      &&/[4(2\theta + 4\cos2\gamma + \cos4\gamma + 1)^2]. \eeqa When
$\max[\frac{2(c_H-c_1)-k}{k-(c_H-c_1)},0]<\cos(2\gamma)$, the
average profits in the iterative game are given by \beqa
\bar{U}^{Q}_1&=&\frac{4[k^2-\Delta^2\theta(1-\theta)]}{8(2+\cos2\gamma)^2}\nonumber\\
&&+\frac{[4k^2+\Delta^2\theta(1-\theta)\cos2\gamma(3+\cos2\gamma)]\cos2\gamma}{8(2+\cos2\gamma)^2},\nonumber\\
\bar{U}^{Q}_2&=&\bar{U}^{Q}_1+\frac{1}{4}\Delta^2\theta(1-\theta).
\eeqa

From the above results, one can find a boundary which separates two
parameter regions A and B labeled by the inequality
$\max[\frac{2(c_H-c_1)-k}{k-(c_H-c_1)},0]<\cos(2\gamma)$ and
$\max[\frac{2(c_H-c_1)-k}{k-(c_H-c_1)},0]\geq\cos(2\gamma)$,
respectively. In the parameter region A, the profits in
Nash-equilibrium in Eq.(24) return to the ones of Ref.\cite{Du2003}
if replacing $\cos2\gamma$ by $\exp[-2\gamma]$ in Ref.\cite{Du2003}.
While in the parameter region B, due to the constraint that the
strategy in Bayes-Nash equilibrium should not exceed the strategy
space $[0,\infty)$. we give out the profits of Bayes-Nash
equilibrium in Eq.(23) which is different with the results in
Ref.\cite{Du2003}.

In Fig.1, the rescaled average profits of the firm 1 and 2 are
plotted as the function of $\Delta/k$ for different values of
$\gamma$. For a fixed value of $\theta$, the degree of asymmetry
$\xi\equiv\frac{\Delta^2(1-\theta)\theta}{k^2}$ is monotonic with
$\Delta/k$. The profit of the firm 2 increases with the degree of
asymmetry, and the profit of the firm 1 decreases with the degree of
asymmetry except for the case with $\gamma=0$, $\theta=0.5$ and
$\Delta/k<\frac{4}{3}$, in which the rescaled profit
$\bar{U}^{Q}_1/k^2$ of the firm 1 keeps fixed. When
$\gamma<\frac{\pi}{4}$, a second-order phase transition-like
behaviors (i.e.
$\frac{\partial{\bar{U}^{Q}_i}/k^2}{\partial{\Delta/k}}$ ($i=1,2$)
is discontinuous) of the rescaled average profits of the firm 1 and
firm 2 in Bayes-Nash equilibrium may be observed as the degree of
asymmetry varies across the boundary of the parameter regions A and
B. In Fig.2, the rescaled average profits of the firm 1 and 2 are
plotted as the function of $\gamma$ for different values of
$\Delta/k$. The asymmetric property of this game can significantly
affect the dependence of the average profits of both firm 1 and 2 on
the degree of "virtual cooperation" $\sin(2\gamma)$. For
$\theta=1/2$ and $0<\Delta/k<\frac{4\sqrt{3}}{9}$, both the profits
of the firm 1 and 2 firstly increase with $\gamma$ and then decrease
with $\gamma$. While for larger asymmetry with $\theta=1/2$ and
$2>\Delta/k>\frac{4\sqrt{3}}{9}$, both the profits of the firm 1 and
2 decrease with $\gamma$. The above results imply that, in this
case, the "virtual cooperation" has advantage role only in the
nearly symmetric or very small asymmetric games. If the degree of
asymmetry exceeds a threshold, the "virtual cooperation" in this
game can suppress the gain of the firm standing on the advantage
side with more information. But for the firm standing on the
disadvantage side due to less information, the "virtual cooperation"
should be regarded as one disaster after another. Similarly, when
$\theta=1/2$ and $\Delta/k>1$, a second-order phase transition-like
behaviors (i.e.
$\frac{\partial{\bar{U}^{Q}_i/k^2}}{\partial{\gamma}}$ ($i=1,2$) is
discontinuous) of the rescaled average profits of the firm 1 and
firm 2 in Nash equilibrium can be observed as the parameter $\gamma$
varies across the boundary of the parameter regions A and B. In
Fig.3, we calculate the total rescaled profit of firm 1 and 2 versus
the parameters $\Delta/k$ or $\gamma$ for $\theta=1/2$. It is found
that in the "ideal virtual cooperation" case with
$\gamma={\frac{\pi}{4}}^-$, the total profit in Nash equilibrium is
invariant against the asymmetry of this game. In other cases, the
total profit always increases with $\Delta/k$. The smaller the
parameter $\gamma$, the more significant influence of the asymmetry
on the total profit.

\begin{figure}
\centerline{\includegraphics[width=2.5in]{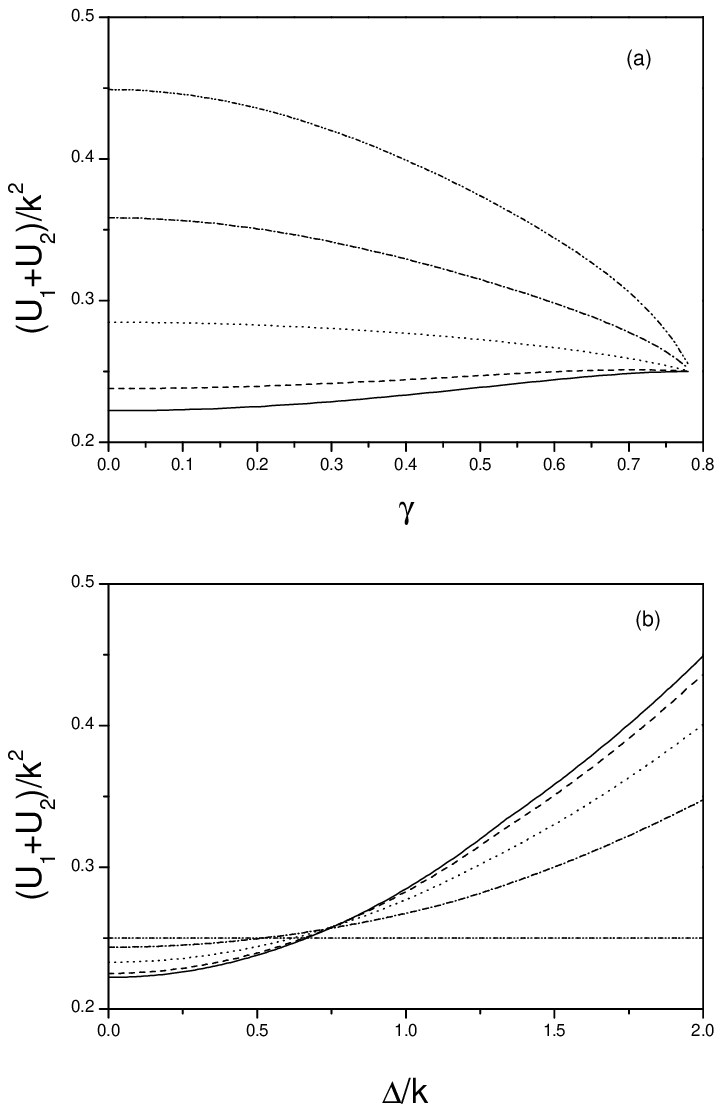}}
\caption{(a) The scaled total payoff $(U_1+U_2)/k^2$ at Nash
equilibrium of two firms is plotted as the function of $\gamma$ for
different values of $\Delta/k$ with $\theta=0.5$. (Solid line)
$\Delta/k=0$; (Dash line) $\Delta/k=0.5$; (Dot line) $\Delta/k=1$;
(Dash dot line) $\Delta/k=1.5$; (Dash dot dot line) $\Delta/k=2^-$.
(b) The scaled total payoff $(U_1+U_2)/k^2$ at Nash equilibrium of
two firms is plotted as the function of $\Delta/k$ for different
values of $\gamma$ with $\theta=0.5$. (Solid line) $\gamma=0$; (Dash
line) $\gamma=\pi/16$; (Dot line) $\gamma=\pi/8$; (Dash dot line)
$\gamma=3\pi/16$; (Dash dot dot line) $\gamma=(\frac{\pi}{4})^{-}$.
}
\end{figure}
\begin{figure}
\centerline{\includegraphics[width=2.5in]{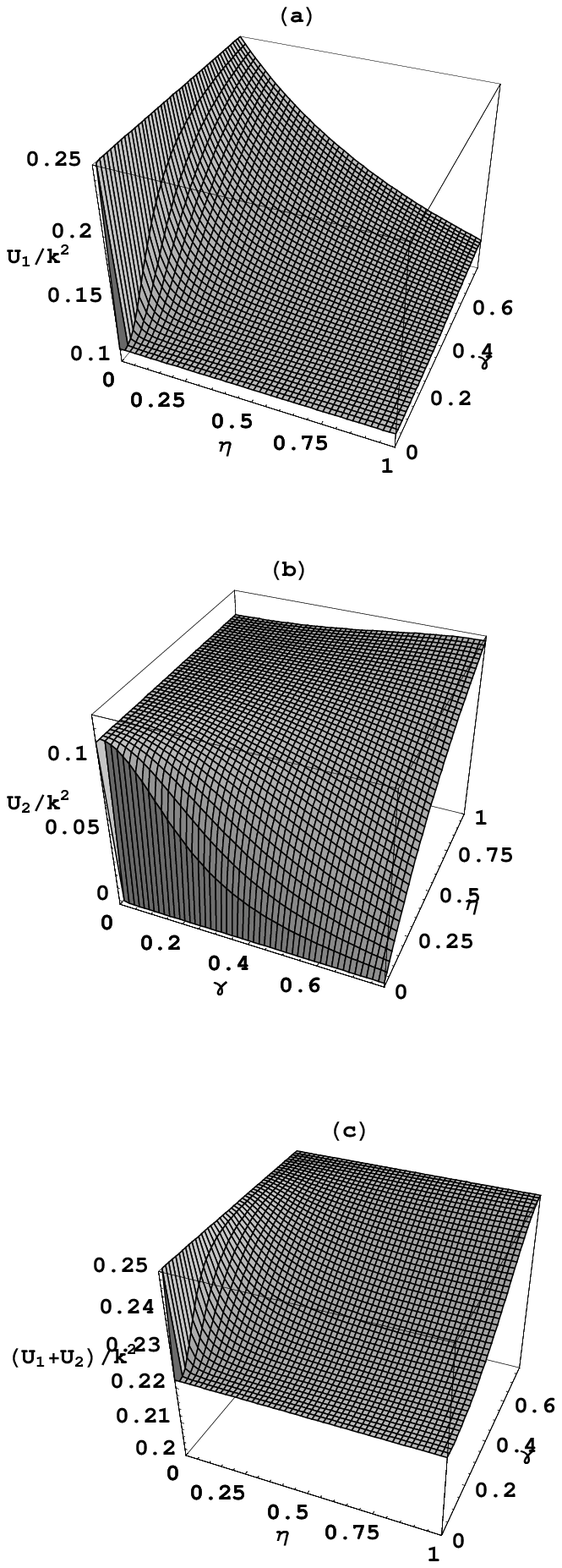}}
\caption{(a) The scaled payoff $U_1/k^2$ of the firm 1, (b) the
scaled payoff $U_2/k^2$ of the firm 2, and (c) the scaled total
payoff $(U_1+U_2)/k^2$ at Nash equilibrium are plotted as the
functions of $\gamma$ and $\eta$.}
\end{figure}

\section * {V. DECOHERENCE-INDUCED ASYMMETRIC QUANTUM GAME}

In the Ref.\cite{Li2006}, by making use of the asymmetrical
entangled states, the quantum model shows some kind of
"encouraging" and "suppressing" effect in profit functions of
different players. Here we discuss how the asymmetric
decoherence can alter the Bayes-Nash equilibrium of the above
quantized scheme of the Cournot's Duopoly, where the asymmetric
decoherence means that two firms experience two different degrees
of decoherence. Let us consider the following specific case in
which the firm 2 encounters a decoherence caused by the photon
loss with the loss rate
$\sqrt{\eta}=e^{-\kappa{\tau}/2}$ ($\tau\in[0,\infty)$) just before the final photon
counting. Obviously, the final measurement gives the respective
quantities of the two firms \beqa
n_1=\frac{1}{2}(x^2_1\cos^2\gamma+x^2_2\sin^2\gamma),\nonumber\\
n_2=\frac{\eta}{2}(x^2_2\cos^2\gamma+x^2_1\sin^2\gamma). \eeqa
Substituting the Eq.(25) into the payoff function, and solving the
two optimization problems yields the Bayes-Nash equilibrium. For
$0<\eta\leq1$ and $0\leq\gamma\leq\frac{\pi}{4}$, the unique
Bayes-Nash equilibrium can be obtained \beqa
x^{\ast{2}}_1=\frac{8k\eta\cos^2\gamma}{1+\eta(6+\eta)+4\eta\cos2\gamma-(1-\eta)^2\cos4\gamma}, \nonumber\\
x^{\ast{2}}_2=\frac{8k\cos^2\gamma}{1+\eta(6+\eta)+4\eta\cos2\gamma-(1-\eta)^2\cos4\gamma}.\eeqa
Meanwhile, the corresponding profits at the Bayes-Nash equilibrium
can be obtained \beqa
{U}^{Q}_1=\xi[1+\eta-(1-\eta)\cos2\gamma],\nonumber\\
{U}^{Q}_2=\xi\eta[1+\eta+(1-\eta)\cos2\gamma], \eeqa where \be
\xi=\frac{2k^2\cos^2\gamma[(1+\eta)^2-(1-\eta)^2\cos^22\gamma]}{[1+\eta(6+\eta)+4\eta\cos2\gamma-(1-\eta)^2\cos4\gamma]^2}\ee
When $\eta\rightarrow0$,
${U}^{Q}_1\rightarrow\frac{k^2}{4+5\delta_{\gamma,0}}$ and
${U}^{Q}_2\rightarrow\frac{k^2\delta_{\gamma,0}}{9}$, where
$\delta_{\gamma,0}$ equals 1 for $\gamma=0$ and is zero elsewhere.
When $\gamma=0$, ${U}^{Q}_1={U}^{Q}_2=\frac{k^2}{9}$, which shows
the payoff does not depend on $\eta$ and implies the Nash
equilibrium of the classical game is robust against the asymmetric
photon-loss.

In Fig.4, the rescaled profits of two firms in the Nash equilibrium
are plotted as the function of $\gamma$ and $\eta$. For
$\gamma\neq0$, the asymmetric decoherence encourages the profit
${U}^{Q}_1$ of the firm 1 and suppresses the profit ${U}^{Q}_2$ of
the firm 2. We can find ${U}^{Q}_1$ and ${U}^{Q}_2$ exhibit the
sharp decline and ascent in the end of asymmetric photon loss for
those cases with very small value of $\gamma\neq0$. In the initial
stage of photon-loss with $\eta>0.5$, "virtual cooperation"
accelerates the encouragement and suppression effects of the
asymmetric photon loss. The asymmetric photon loss plays a role in
transferring the profit from the firm 2 to firm 1. Surprisingly, the
asymmetric photon loss can improve the total profits of the firm 1
and 2 in the Nash equilibrium in the situations with
$0<\gamma<\frac{\pi}{4}$. In the ideal "virtual cooperation", i.e.
$\gamma=\frac{\pi}{4}$, the total profits are kept fixed against the
asymmetric photon loss.

\section * {VI. CONCLUSIONS}

In this paper, we present a simulation scheme of the continuous
variable quantized Cournot's duopoly, in which not any intermediate
quantum entanglement has been involved. The influence of measuring
apparatus, symmetric or asymmetric photon loss, and asymmetric
information on their Nash equilibria has been investigated. It is
shown that the scheme using classical measuring apparatus is
advantage to the one using the quantum measuring apparatus. Being
different from the previous quantized Cournot's duopoly involving
entanglement, this simulation scheme is also symmetric photon loss
free; While for asymmetric photon loss, the profits in Nash
equilibrium exhibit a transfer from one firm to the other.
Simultaneously, the total profit in Nash equilibrium increases with
the asymmetric photon loss except for two extreme cases, i.e. the
complete no "virtual cooperation" case and the ideal "virtual
cooperation" case.

In the cases with asymmetric information, a second-order phase
transition-like behavior of the average profits of the firm 1 and firm 2 in Nash
equilibrium can be observed as the degree of asymmetry or
the degree of "virtual cooperation" vary. The "virtual
cooperation" has advantage role for total profit in Nash
equilibrium only in the nearly symmetric or very small asymmetric
games. If the degree of asymmetry exceeds a threshold value, the "virtual
cooperation" in this game can suppress the gain of the firm standing
on the advantage side with more information. But for the firm
standing on the disadvantage side due to less information, the
"virtual cooperation" should be regarded as one disaster after
another. For the total profit, it is found that, in the "ideal
virtual cooperation" case with $\gamma={\frac{\pi}{4}}^-$, the total
profit in Nash equilibrium is invariant against the asymmetry of
this game. In other cases with $\gamma<\frac{\pi}{4}$, the total
profit always increases with $\Delta/k$. The smaller the parameter
$\gamma$, the more significant influence of the asymmetry on the
total profit.

\bibliographystyle{apsrev}

\end{document}